\documentclass[%
aps,
twocolumn,
superscriptaddress,
amsmath,amssymb,
prl,
floatfix,
]{revtex4-2}
\usepackage{url}
\usepackage{graphicx}
\usepackage{placeins}
\usepackage{dcolumn}
\usepackage{bm}
\usepackage[colorlinks, linkcolor=red, anchorcolor=green, citecolor=blue]{hyperref}

\usepackage{physics}
\usepackage{dsfont}
\usepackage{tikz,pgfplots}
\usepackage{enumerate}
\usepackage{subfigure}
\usepackage{bbold}
\usepackage{makecell}
\usepackage{array}
\makeatletter
\newcommand{\thickhline}{%
    \noalign {\ifnum 0=`}\fi \hrule height 1pt
    \futurelet \reserved@a \@xhline
}
\newcolumntype{"}{@{\hskip\tabcolsep\vrule width 1pt\hskip\tabcolsep}}
\makeatother
\usepackage{multirow}
\usepackage{tabu}
\usepackage{textcomp,booktabs}
\usepackage{colortbl}
\usepackage[T1]{fontenc}
\definecolor{mygray}{gray}{0.9}
\definecolor{mypink}{rgb}{0.99,0.91,0.95}
\definecolor{mycyan}{cmyk}{0.3,0,0,0}

\newcommand{\mB}{\mathcal{B}}

\newcommand{\mE}{\mathcal{E}}

\newcommand{\mI}{\mathcal{I}}

\newcommand{\mO}{\mathcal{O}}

\newcommand{\mU}{\mathcal{U}}
\newcommand{\T}{\mathbf{T}}

\newcommand{\1}{\mathbb{1}}

\makeatletter
\newcommand*{\rom}[1]{\expandafter\@slowromancap\romannumeral #1@}
\makeatother

\usepackage{amsthm}
\newtheorem*{thm*}{Theorem}
\newtheorem{thm}{Theorem}
\newtheorem{lem}{Lemma}

\usepackage{tcolorbox}
\tcbuselibrary{theorems}

\newtcbtheorem[number within=section, use counter*=thm]{mythm}{Theorem}%
{colback=red!5,colframe=red!35!red,fonttitle=\bfseries}{thm}

\usepackage{tcolorbox}
\tcbuselibrary{theorems}

\newtcbtheorem[number within=section, use counter*=thm]{mylem}{Lemma}%
{colback=green!5,colframe=green!35!black,fonttitle=\bfseries}{lem}

\usepackage{tcolorbox}
\tcbuselibrary{theorems}

\newtcbtheorem[number within=section, use counter*=thm]{mydef}{Definition}%
{colback=yellow!5,colframe=yellow!70!black,fonttitle=\bfseries}{def}

\usepackage{tcolorbox}
\tcbuselibrary{theorems}

\newtcbtheorem[number within=section, use counter*=thm]{mycor}{Corollary}%
{colback=blue!5,colframe=blue!70!black,fonttitle=\bfseries}{cor}

\usepackage{tcolorbox}
\tcbuselibrary{theorems}

\newtcbtheorem[number within=section, use counter*=thm]{myprop}{Proposition}%
{colback=pink!5,colframe=pink!70!black,fonttitle=\bfseries}{prop}

\pgfplotsset{compat=1.17}
\begin{document}


\title{No Practical Quantum Broadcasting: Even Virtually}

\author{Yunlong Xiao}
\email{mathxiao123@gmail.com}
\affiliation{Quantum Innovation Centre (Q.InC), Agency for Science Technology and Research (A*STAR), 2 Fusionopolis Way, Innovis \#08-03, Singapore 138634, Republic of Singapore}
\affiliation{Institute of High Performance Computing (IHPC), Agency for Science, Technology and Research (A*STAR), 1 Fusionopolis Way, \#16-16 Connexis, Singapore 138632, Republic of Singapore}

\author{Xiangjing Liu}
\affiliation{Nanyang Quantum Hub, 
School of Physical and Mathematical Sciences,
Nanyang Technological University, 
Singapore 637371, 
Republic of Singapore}
\affiliation{CNRS@CREATE, 1 Create Way, 08-01 Create Tower, Singapore 138602, Singapore}
\affiliation{MajuLab, CNRS-UCA-SU-NUS-NTU International Joint Research Laboratory}
\affiliation{Centre for Quantum Technologies, National University of Singapore, Singapore 117543, Singapore}

\author{Zhenhuan Liu}
\email{liu-zh20@mails.tsinghua.edu.cn}
\affiliation{Center for Quantum Information, Institute for Interdisciplinary Information Sciences, Tsinghua University, Beijing 100084, China}

             
\begin{abstract} 
Quantum information cannot be broadcast -- an intrinsic limitation imposed by quantum mechanics. 
However, recent advances in virtual operations offer new insights into the no-broadcasting theorem. 
Here, we focus on the practical utility and introduce sample efficiency as a fundamental constraint, requiring any practical broadcasting protocol perform no worse than the na\"ive approach of direct preparation and distribution.
We prove that no linear process -- whether quantum or beyond -- can simultaneously uphold sample efficiency, unitary covariance, permutation invariance, and classical consistency. 
This leads to a \emph{no-practical-broadcasting theorem}, which places strict limits on the practical distribution of quantum information.
By applying Schur-Weyl duality, we establish the uniqueness of the canonical 1-to-$N$ virtual broadcasting map that satisfies the latter three conditions, provide its construction, and determine its sample complexity through semidefinite programming.
Finally, we explore the interplay between virtual broadcasting and a quantum spacetime framework, known as the pseudo-density operator, showing that their correspondence holds only in the 1-to-2 case, underscoring the fundamental asymmetry between spatial and temporal statistics in the quantum world.
\end{abstract}

\maketitle

\begin{figure}[ht]
\centering   
\includegraphics[width=0.48\textwidth]{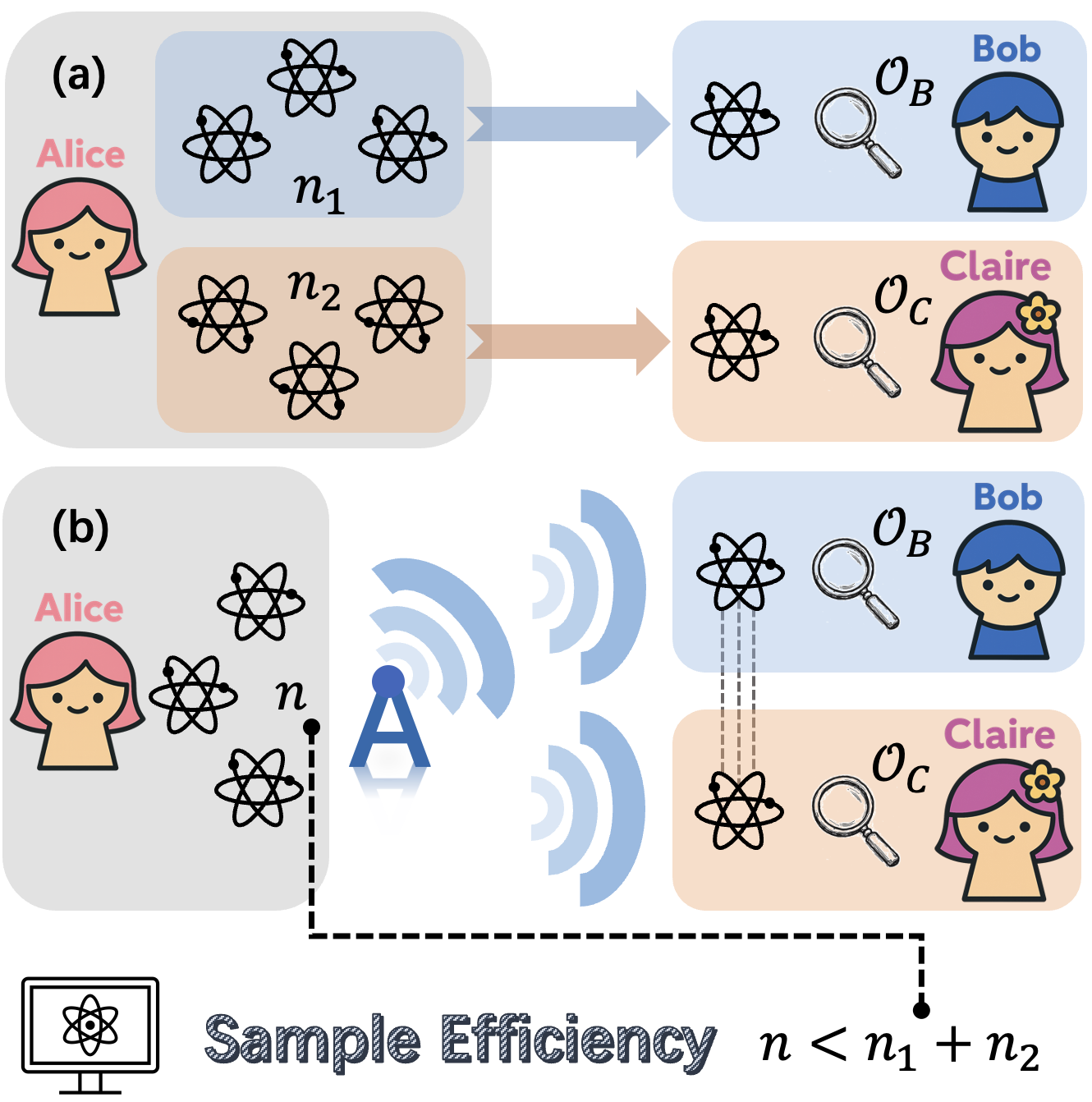}
\caption{(Color online) \textbf{Practical Broadcasting}.  
(a) Na\"ive Protocol: Consider a scenario where Bob requires 
$n_1$ copies and Claire needs $n_2$ copies of a quantum state to achieve their respective target accuracies. In the na\"ive broadcasting strategy, Alice prepares $n_1+n_2$ independent copies of the state and directly distributes $n_1$ to Bob and $n_2$ to Claire.
(b) Sample Efficient Protocol: A broadcasting protocol is deemed sample efficient if Alice can enable both Bob and Claire to achieve their respective accuracy requirements using a total of only $n$ copies of the state, where $n<n_1+n_2$.
}
\label{fig:Sample_Efficiency}
\end{figure}

\noindent \textbf{Introduction}--Quantum theory unlocks transformative capabilities in information processing, underpinning applications in quantum communication~\cite{bennett1984quantum} and computation~\cite{shor1997factoring}.
However, encoding information in quantum systems is constrained by physical laws. For instance, the no-cloning theorem demonstrates that no process can perfectly replicate an arbitrary quantum state~\cite{Wootters1982clone,DIEKS1982clone}. This restriction extends beyond exact duplication: even when only the marginal statistics of the output need to match the original state -- known as broadcasting (see Fig.~\ref{fig:Sample_Efficiency}) -- it remains inherently prohibited~\cite{barnum1996broadcasting}. This intricate balance between quantum advantages and inherent limitations defines the operational landscape of the quantum world, shaping both its potential and constraints.

Although no-go theorems seem to establish definitive limits within quantum theory, a path forward exists. Recent advancements demonstrate that by combining quantum processes with classical post-processing, virtual operations transcend conventional quantum manipulations, driving progress in virtual cooling~\cite{cotler2019cooling}, quantum error mitigation~\cite{temmeErrorMitigationShortDepth2017,endoPracticalQuantumError2018}, and resource distillation~\cite{yuan2024virtual}. 
This raises the questions: Can virtual operations circumvent no-broadcasting~\cite{PhysRevLett.132.110203,PhysRevA.110.012458,8vrg-tvsd} while achieving better performance in sample complexity compared to the na\"ive approach of direct preparation and distribution? 
If the transition from quantum operations to virtual operations yields substantial benefits, could even more remarkable capabilities be gained by fully exploiting the linearity of quantum mechanics, extending beyond virtual operations to more general linear maps?

In this work, we prove that no practical broadcasting map exists among linear maps that satisfy the conditions of sample efficiency, unitary covariance, permutation invariance, and classical consistency. 
As a starting point, we characterize the class of linear maps obeying the latter three conditions and identify a unique 1-to-$N$ virtual broadcasting map that fulfills them. This generalizes previous results~\cite{PhysRevLett.132.110203}, which were limited to the 1-to-2 case, and sets the stage for a broader analysis of virtual broadcasting in quantum mechanics. We then explore its connection to pseudo-density operators~\cite{fitzsimons2015quantum}, a framework that extends conventional density matrices to incorporate temporal correlations, and uncover a fundamental divergence between the two. Finally, we show that achieving the desired accuracy in virtual broadcasting necessitates a significantly larger number of samples than the na\"ive approach, where copies are merely prepared and sent to each receiver. This discrepancy not only undermines sample efficiency but also highlights the impracticality of broadcasting under the linearity assumption in quantum mechanics.

\noindent \textbf{Practical Broadcasting}--To make quantum broadcasting operationally meaningful, it is essential to understand its sample complexity, which gives rise to the fundamental constraint of {\it sample efficiency} (SE). For clarity, we begin with the simplest nontrivial case: broadcasting to two receivers. As illustrated in Fig.~\ref{fig:Sample_Efficiency}, such a process involves three parties: The sender, Alice, aims to distribute an unknown quantum state $\rho$ to two receivers, Bob and Claire, such that their local measurements reproduce the same statistics as those of the original state. Let us assume that Bob and Claire wish to measure different observables, $\mO_B$ and $\mO_C$, requiring $n_1$ and $n_2$ copies of $\rho$ to achieve the desired accuracy, respectively. A practical broadcasting process should complete this task using strictly fewer than $n_1+n_2$ copies (see Fig.~\ref{fig:Sample_Efficiency}(b)).
Otherwise, sending $n_1$ copies to Bob and $n_2$ copies to Claire would trivialize the task (see Fig.~\ref{fig:Sample_Efficiency}(a)). The analysis naturally extends to the one-to-many setting: if multiple receivers each require $n_i$ copies to estimate their respective observables $\mO_i$ to a given precision, a non-trivial broadcasting process must accomplish all tasks using strictly fewer than $\sum_i n_i$ copies. This defines our operational criterion for SE, ensuring that broadcasting is not merely a quantum data redistribution, but a genuine process with non-trivial quantum resource advantages.

Beyond SE, a broadcasting map should satisfy three additional fundamental conditions: {\it unitary covariance} (UC), {\it permutation invariance} (PI), and {\it classical consistency} (CC). UC dictates that applying a unitary operation to the sender's system prior to broadcasting is operationally equivalent to first broadcasting and subsequently applying the same unitary to each receiver's system. PI ensures that any permutation of the receivers post-broadcasting leaves the output unchanged, thereby preserving the inherent symmetry of the distribution process. CC asserts that, in the limit of complete dephasing across all quantum systems, the broadcasting map must reduce to a classical broadcasting process between the sender and receivers. A linear map that satisfies all four conditions -- SE, UC, PI, and CC -- is referred to as a {\it practical broadcasting map}.


\noindent \textbf{Canonical Form}--We begin by defining UC, PI, and CC mathematically, and show that these conditions uniquely determine a map -- the canonical 1-to-$N$ virtual broadcasting map -- within the set of linear maps. 
Notably, we make no assumptions about the map being {\it Hermitian-preserving}, {\it trace-preserving}, or even {\it broadcasting}. Our focus is strictly on linear maps, where UC, PI, and CC are the only conditions considered.
A map $\mE: A \to B_1\cdots B_N$ is said to be broadcasting if and only if $\Tr_{\overline{B}_i}\circ\,\mE=\mI_{A\to B_i}$. Here, $\Tr_{\overline{B}_i}$ denotes the partial trace over all receiver subsystems except $B_i$, and $\mI$ denotes the identity channel.

The UC condition requires that for any unitary gate $\mU$, the map $\mE$ satisfies
\begin{align}\label{eq:Condition1}
    \underbrace{\mU\otimes\cdots\otimes\mU}_\text{$N$}
    \circ\,\mE = \mE\circ\mU,
\end{align}
where the notation $\circ$ represents the composition of maps. The PI condition is given by
\begin{align}\label{eq:Condition2}
    P_{\pi}\circ\mE &= \mE,\quad
    \forall\,\pi\in \mathfrak{S}_N,
\end{align}
where the permutation map $P_{\pi}$ is defined by $P_{\pi}(X) := V_{\pi} X V_{\pi}^{\T}$, with $V_{\pi}$ being the matrix representation of $\pi$ acting on the output systems $B_1\cdots B_N$, and the superscript $\T$ denoting matrix transposition.
Finally, the CC condition is expressed as
\begin{align}\label{eq:Condition3}
    \underbrace{\Delta\otimes\cdots\otimes\Delta}_\text{$N$}
    \circ\,\mE\circ\Delta
    &=\mB_{\text{N-cl}},
\end{align}
where $\Delta(\cdot):=\sum_{i}\bra{i}\cdot\ket{i}\ketbra{i}{i}$ represents the completely dephasing channel in some basis $\{\ket{i}\}$, and $\mB_{\text{N-cl}}(\ketbra{i}{j}):=\delta_{ij}\ketbra{i}{i}\otimes\cdots\otimes\ketbra{i}{i}$ is the classical 1-to-$N$ broadcasting map.
The following theorem establishes the existence of a unique linear map that satisfies these constraints.

\begin{thm}[Canonical 1-to-$N$ Virtual Broadcasting]\label{thm:1-to-$N$}
Any linear map $\mE: A \to B_1\cdots B_N$ that satisfies UC, PI, and CC must coincide with $\mB_N$
\begin{align}\label{eq:B_N_anti}
    \mB_N(\rho):=
    \frac{1}{2}
    \{\rho_{\mathbb{N}},V_{\mathbb{N}}\}.
\end{align}
Here, the average state $\rho_{\mathbb{N}}$ is given by
\begin{align}\label{eq:rho_bbN}
    \rho_{\mathbb{N}}:=\frac{1}{N}\sum_{i=1}^N
    \underbrace{\1_{B_1}\otimes\cdots\otimes\1_{B_{i-1}}}_\text{$i-1$}
    \otimes\,\rho\otimes
    \underbrace{\1_{B_{i+1}}\otimes\cdots\otimes\1_{B_N}}_\text{$N-i$},
\end{align}
and $V_{\mathbb{N}}$ is defined as the average over all permutation matrices corresponding to N-cycles within the symmetric group $\mathfrak{S}_{N}$ 
\begin{align}\label{eq:V_bbN}
    V_{\mathbb{N}}
    :=
    \frac{1}{(N-1)!}
    \sum_{\substack{\pi: \text{N-cycle} \\ \pi\in\mathfrak{S}_{N}}}
    V_{\pi}.
\end{align}
\end{thm}

Our proof, grounded in Schur-Weyl duality~\cite{Collins2006random} and tensor network techniques, is detailed in~\cite{8g6j-w7ld}. We make two remarks: First, our result shows that the uniqueness of virtual broadcasting under the constraints of UC, PI, and CC holds broadly for general 1-to-$N$ linear maps. When $N=2$, it encompasses the main finding of~\cite{PhysRevLett.132.110203}. Second, although we do not assume the properties of Hermitian-preserving, trace-preserving, or broadcasting, the canonical broadcasting map $\mB_N$ automatically satisfies all of them. Particularly, our proof shows that UC and CC are sufficient to imply broadcasting, while the inclusion of the PI further ensures Hermitian preservation. This is counterintuitive, as we explore broadcasting without assuming the broadcasting condition upfront. Moreover, the broadcasting condition inherently guarantees trace-preserving, obviating the need for this assumption. This highlights that the virtual broadcasting arises naturally from the structural constraints rather than being an a priori assumption. Even in the 1-to-2 case, our approach sets itself apart from~\cite{PhysRevLett.132.110203}, where the authors found that Hermitian-preserving could be derived as a result rather than an assumption. Here, we go further -- demonstrating that even the broadcasting condition does not need to be assumed.

Beyond broadcasting, $\mB_2$ also serves as fundamental structural bridge between spatial and temporal aspects of quantum theory. Mathematically, it coincides with the pseudo-density operator of $\rho$ under an identity channel $\mI$~\cite{Fitzsimons2015}, captures the real part of the two-point correlator for qudit systems~\cite{buscemi2013directobservationtwopointquantum,doi:10.1142/S0219749915600023}, and aligns with the quantum state-over-time~\cite{horsman2017qsot,fullwood2022qsot,PRXQuantum.4.020334,PhysRevResearch.6.033144,lie2024uniquemultipartiteextensionquantum}. Physically, the permutation invariance of $\mB_2$ -- typically regarded as a spatial symmetry -- admits a reinterpretation as time-reversal symmetry when viewed through the lens of the pseudo-density operator formalism. This reframing positions $\mB_2$ as a form of temporal broadcasting, suggesting that spatially defined quantum correlations may possess consistent temporal counterparts. Since both the pseudo-density operator and quantum state-over-time frameworks naturally extend to multiple time points, a key open question is whether multi-receiver virtual broadcasting maps preserve this spatiotemporal correspondence. To address this, we examine the structure of 1-to-3 broadcasting $\mB_3$ through tensor networks
\begin{widetext}
\begin{align}
    \mB_3(\rho)
    =
    \frac{1}{6}
    \left(
    \raisebox{-1.8ex}{\includegraphics[height=2.5em]{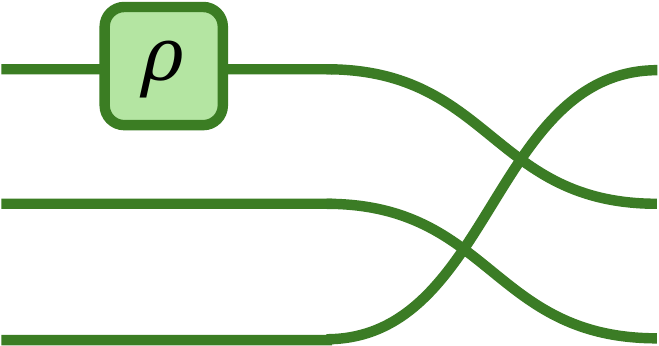}}
    +
    \raisebox{-1.8ex}{\includegraphics[height=2.1em]{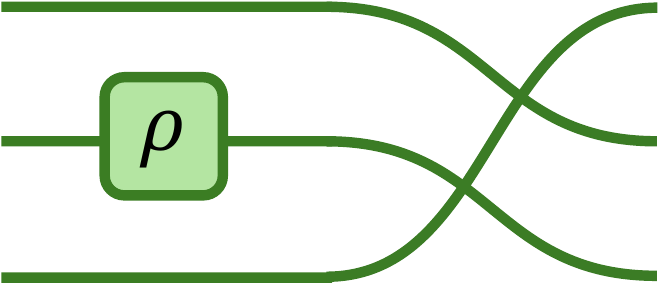}}
    +
    \raisebox{-2.8ex}{\includegraphics[height=2.5em]{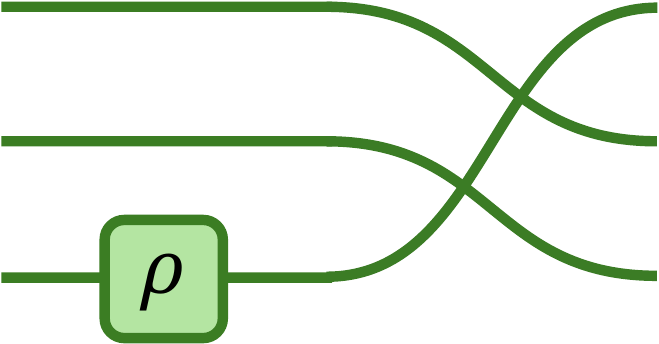}}
    +
    \raisebox{-1.8ex}{\includegraphics[height=2.5em]{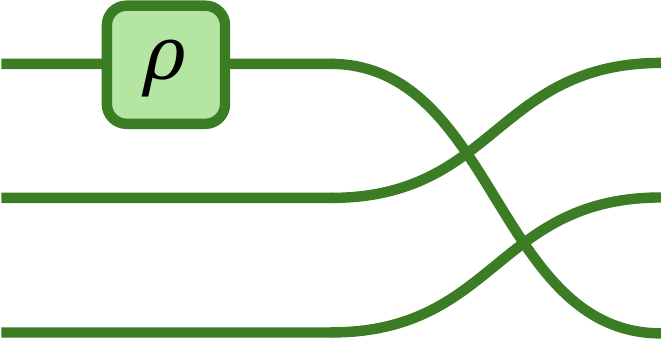}}
    +
    \raisebox{-1.8ex}{\includegraphics[height=2.1em]{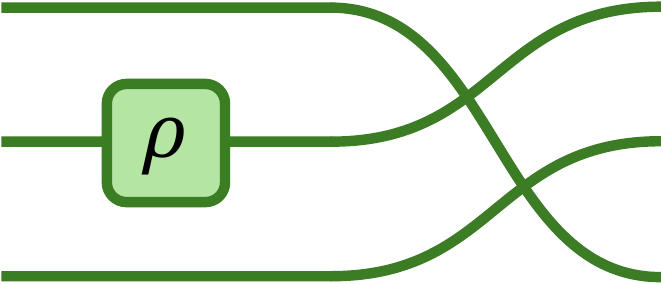}}
    +
    \raisebox{-2.8ex}{\includegraphics[height=2.5em]{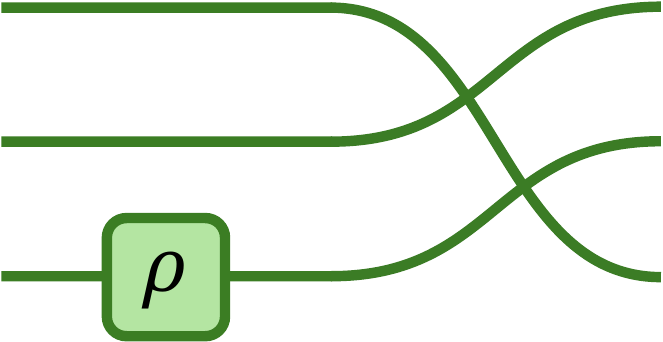}}
    \right).\label{eq:B_3}
\end{align}
\end{widetext}
In this diagram, the box with two legs symbolizes the density matrix $\rho$, the straight line represents the identity operator $\1$, and the crossed lines indicate permutations of the systems.
On the other hand, the pseudo-density operator corresponding to an initial state $\rho$, evaluated at three time points and evolved through two successive identity channels, takes the form~\cite{liu2023quantum}
\begin{align}
    &\frac{1}{4}
    \left(
    \raisebox{-1.8ex}{\includegraphics[height=2.5em]{B_3_1.pdf}}
    +
    \raisebox{-2.7ex}{\includegraphics[height=2.4em]{B_3_3.pdf}}+
    \raisebox{-1.7ex}{\includegraphics[height=2.4em]{B_3_4.pdf}}
    +
    \raisebox{-1.6ex}{\includegraphics[height=1.9em]{B_3_5.pdf}}
    \right).\label{eq:R_3_2}
\end{align}
This expression, however, departs from $\mB_3$, revealing that the spatiotemporal correspondence evident in the 1-to-2 case fails to extend to scenarios involving multiple receivers or time points. The discrepancy highlights a deeper structural incompatibility, reflecting a fundamental disjunction between the physical principles governing temporal and spatial correlations. While our companion work offers a more detailed exploration of this direction~\cite{8g6j-w7ld}, a comprehensive investigation remains an open and promising avenue.


\noindent \textbf{Sample Complexity}--Since the canonical virtual broadcasting map $\mB_N$ is the only linear map that satisfies UC, PI, and CC, the question of whether a linear map can also satisfy SE reduces to analyzing the samples required for its realization. 
To frame broadcasting in a practical context, consider a scenario where each receiver $i$ aims to estimate the expectation value of a local observable $\mO_i$. The receiver is said to succeed with confidence if the probability that the estimation error exceeds a threshold $\epsilon_i$ is less than $\delta_i$. By Hoeffding’s inequality~\cite{Hoeffding01031963}, the minimum number of independent copies $n_i$ required to ensure this performance is bounded by $(c^2/2\epsilon_i^2)\ln(2/\delta_i)$, where $c$ is a constant that bounds the difference between measurement outcomes~\cite{8g6j-w7ld}. In this case, satisfying SE requires that the total number of copies used to realize virtual broadcasting be strictly less than $\sum_i n_i$.
If this condition is not met, the sender can instead prepare a total of $\sum_i n_i$ copies of the state $\rho$, distributing $n_i$ copies to the $i$-th receiver. This direct allocation allows each receiver to perform their task independently, with greater efficiency than using $\mB_N$, thereby rendering the process of broadcasting quantum information trivial in practice.

Simulating $\mB_N$ requires samples determined by its decomposition into linear combinations of quantum channels, i.e., completely positive and trace-preserving maps. 
Given that $\mB_N$ can be expressed as $a\,\mE_1-b\,\mE_2$, where $\mE_1$ and $\mE_2$ are quantum channels and $a, b$ are non-negative numbers, the minimum number of samples required for simulation is given by $(a+b)^2 n_{\text{Q}}$, with $n_{\text{Q}}:=\max_i n_i$ being the sample complexity of a ``physical'' broadcasting.
Therefore, optimizing over all possible decompositions leads to the minimal samples for realizing $\mB_N$, which is characterized by the following semidefinite programming~\cite{doi:10.1137/1038003}, 
\begin{align}\label{eq:SDP}
    u_N:=\min \quad & a+b\notag\\
    \text{s.t.} \quad & J_1-J_2= J^{\mB_N},\,\, J_1\geqslant0,\,\, J_2\geqslant0,\notag\\
    &\Tr_{B_1\cdots B_N}[J_1]=a\,\1_A,\,\,
    \Tr_{B_1\cdots B_N}[J_2]=b\,\1_A,
\end{align}
where $J^{\mB_N}$ denotes the Choi operator~\cite{CHOI1975285,JAMIOLKOWSKI1972275} of $\mB_N$. 

We now proceed to compare the sample requirements for virtual broadcasting, $n_{\text{V}}:=u_N^2 n_{\text{Q}}$, with that of the na\"ive approach, $\sum_i n_i$. In the 1-to-2 scenario, the canonical form is given by $\mB_2(\rho)=\{\rho\otimes\1, S\}/2$, with $\{\cdot,\cdot\}$ denoting the anti-commutator, $S$ the SWAP operator, and $\1$ the identity operator. Under this configuration, the na\"ive approach requires a total of $n_1+n_2$ copies, and the scaling factor $u_2$ in virtual broadcasting admits a closed-form expression captured in the lemma below.

\begin{lem}
For the canonical 1-to-2 virtual broadcasting map $\mB_2$, the sampling overhead is determined by the system dimension $d := \dim A$, leading to $u_2=d$. The decomposition into quantum channels provided below offers an experimentally implementable realization, while minimizing sample overhead
\begin{align}
    \mB_2=\frac{d+1}{2}\mE_1-\frac{d-1}{2}\mE_2,
\end{align}
where the Choi operators of $\mE_1$ and $\mE_2$ are given by
\begin{align}
    J^{\mE_1}
    &=
    \frac{
    \raisebox{-1.6ex}{\includegraphics[height=1.8em]{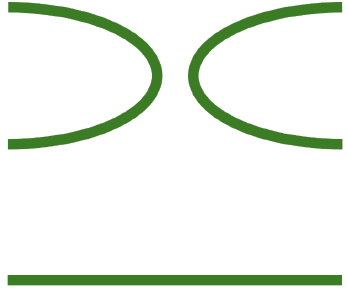}}
    +
    \raisebox{-1.6ex}{\includegraphics[height=1.8em]{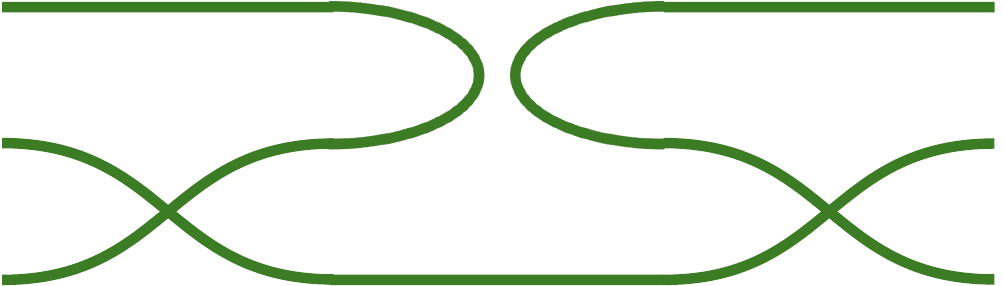}}
    -
    \raisebox{-1.6ex}{\includegraphics[height=1.8em]{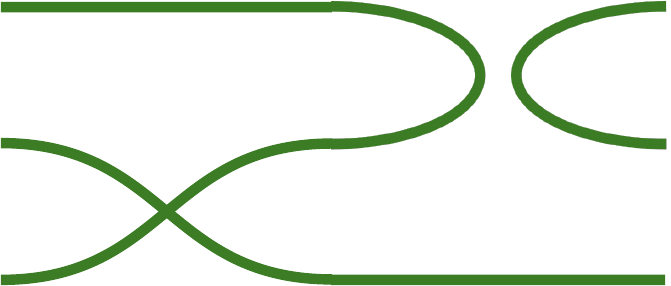}}
    -
    \raisebox{-1.6ex}{\includegraphics[height=1.8em]{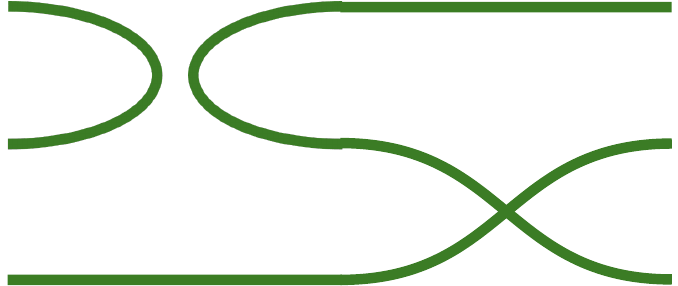}}
    }{2(d-1)},\\
    J^{\mE_2}
    &=
    \frac{
    \raisebox{-1.6ex}{\includegraphics[height=1.8em]{M.pdf}}
    +
    \raisebox{-1.6ex}{\includegraphics[height=1.8em]{NMN.pdf}}
    +
    \raisebox{-1.6ex}{\includegraphics[height=1.8em]{MN.pdf}}
    +
    \raisebox{-1.6ex}{\includegraphics[height=1.8em]{NM.pdf}}
    }{2(d+1)}.
\end{align}
\end{lem}

\begin{figure}[t]
    \centering   
    \includegraphics[width=0.48\textwidth]{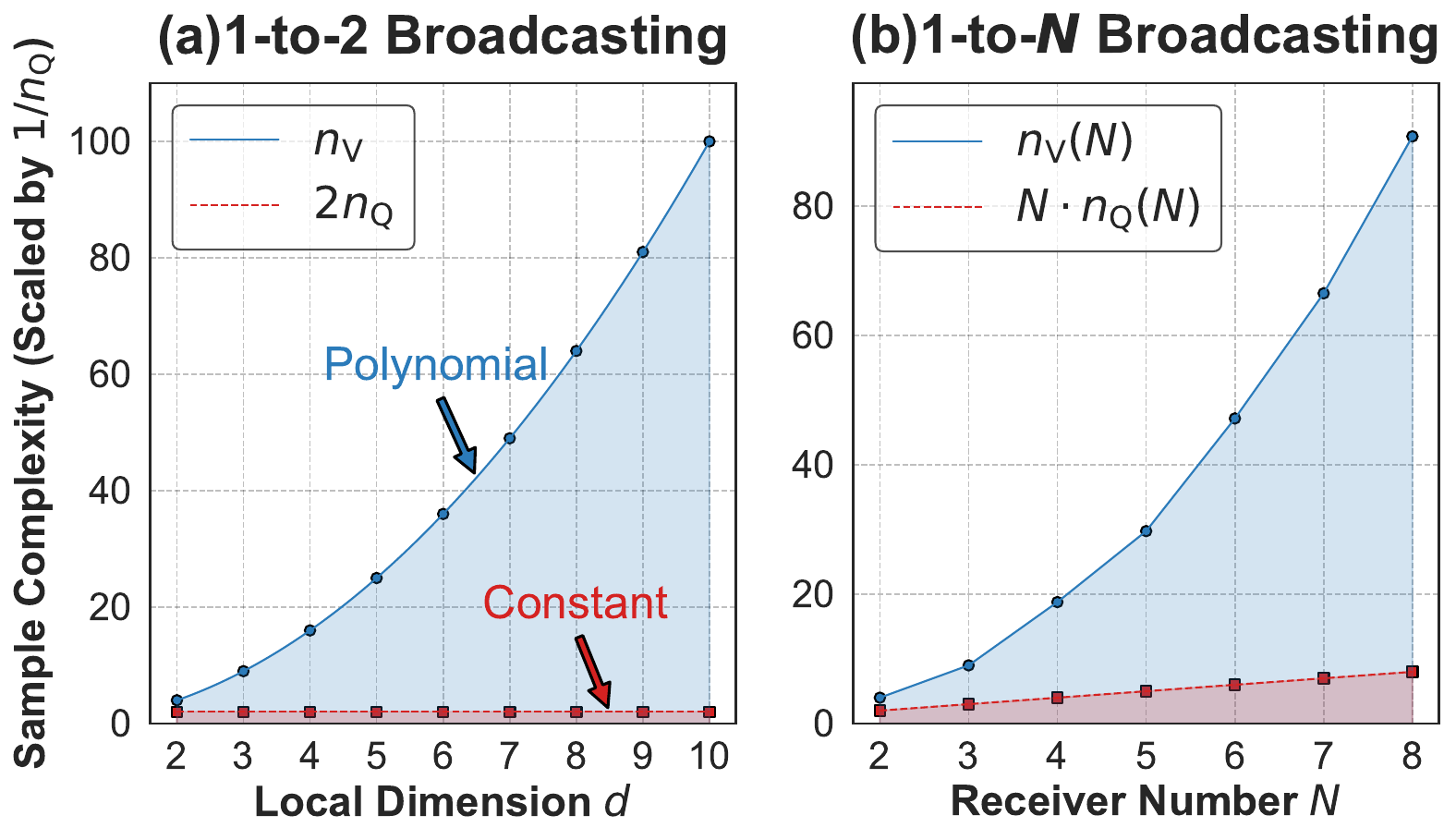}
    \caption{(Color online) \textbf{Comparing Sample Complexity}.
    The vertical axis represents the sample complexity, scaled by $1/n_{\text{Q}}$. In (a), the horizontal axis corresponds to the dimension $d$ of the state being broadcast, while in (b), it denotes the number $N$ of receivers, with $d=2$. The blue line illustrates the sample complexity required to implement the canonical virtual broadcasting map -- specifically, $\mB_2$ in (a) and $\mB_N$ in (b). In both panels, the red dashed line indicates the number of copies needed for the na\"ive approach, where i.i.d. copies of $\rho$ are prepared and sent directly to the receivers.
   }
    \label{fig:SC}
\end{figure}

\noindent
Here, the semi-circle represents a maximally entangled state, while the crossed lines denote the SWAP operator.
A detailed proof of the lemma is available in~\cite{8g6j-w7ld}. In quantum theory, the system's dimension $d$ must be at least $2$, implying
\begin{align}
    n_{\text{V}}
    = d^2 n_{\text{Q}}
    > 2n_{\text{Q}}
    \geqslant n_1+n_2.
\end{align}
In other words, to achieve the desired confidence level for each receiver's estimation, a higher sample complexity is required for simulating $\mB_2$ compared to the na\"ive approach (see Fig.~\ref{fig:SC}(a)). Consequently, the canonical virtual broadcasting map fails to meet the SE condition. 

\begin{thm}[No Practical Quantum Broadcasting]\label{thm:no-broadcasting}
    A 1-to-2 linear map that fulfills the conditions of SE, UC, PI, and CC does not exist.
\end{thm}

We proceed to analyze the sample complexity of simulating the canonical 1-to-$N$ virtual broadcasting protocol and compare it with the na\"ive strategy, which requires $\sum_i n_i$ copies, as illustrated in Fig.~\ref{fig:SC}(b). The comparison demonstrates that virtual broadcasting incurs a higher cost, highlighting its impracticality.

\noindent \textbf{Discussions}--In this work, we investigate whether a linear map can satisfy the conditions of sample efficiency, unitary covariance, permutation invariance, and classical consistency simultaneously, and conclude negatively. 
We first establish that a unique linear map, the canonical 1-to-$N$ virtual broadcasting map, satisfies the latter three conditions, generalizing previous work on 1-to-2 virtual broadcasting. 
This framework allows us to examine the connection between broadcasting, which encodes quantum correlations in space, and pseudo-density operators, which capture quantum correlations in time. 
Our findings reveal that the apparent correspondence between 1-to-2 virtual broadcasting and pseudo-density operators at two time points is purely coincidental.
The second contribution of this work involves analyzing the number of samples required for implementing virtual broadcasting, where we show that if only classical statistical data (rather than the quantum state) needs to be matched, distributing a sufficient number of copies to each receiver can achieve the task more efficiently, using fewer copies than the virtual broadcasting. 

Our results reaffirm the no-broadcasting theorem as a foundational limitation in quantum information theory. Even when the set of quantum system manipulations is extended from quantum channels to all linear maps, broadcasting quantum information remains fundamentally infeasible in practice. Although previous studies have suggested that virtual operations might offer a potential solution for simulating the statistical outcomes of broadcasting, we have shown that this is not the case. The significantly higher sample overhead required for such tasks, relative to the na\"ive protocol, makes them impractical for real-world implementation.
This insight naturally raises open questions: Which no-go theorems can be overcome by broadening the set of allowed operations, and which remain fundamentally insurmountable? 
Furthermore, for the latter, is there a unifying principle that governs their persistence? 
We leave these intriguing questions for future exploration.


\section*{Acknowledgments}
Yunlong Xiao gratefully acknowledges Sandu Popescu, Giulio Chiribella, Francesco Buscemi, and Mio Murao for stimulating discussions during QMQI 2024 in Okinawa. Appreciation is also extended to Yuxiang Yang, Honghao Fu, Penghui Yao, and Naihuan Jing for insightful conversations at QIP 2025 in Raleigh. Furthermore, he thanks Jian Zhang, Ryuji Takagi, Varun Narasimhachar, Davit Aghamalyan, Ximing Wang, Matthew Simon Tan, Benchi Zhao, Jayne Thompson, Kishor Bharti, Seok Hyung Lie, and Jun Ye for valuable discussions.  
This research is supported by A*STAR under C23091703 and Q.InC Strategic Research and Translational Thrust.
Yunlong Xiao acknowledges support from A*STAR under its Central Research Funds and Career Development Fund (C243512002).
Xiangjing Liu is funded by the National Research Foundation, Prime Minister’s Office, Singapore under its Campus for Research Excellence and Technological Enterprise (CREATE) programme. 
Zhenhuan Liu acknowledges the support of the National Natural Science Foundation of China Grant No.~12174216 and the Innovation Program for Quantum Science and Technology Grant No.~2021ZD0300804 and No.~2021ZD0300702. 



\begin{thebibliography}{28}%
	\makeatletter
	\providecommand \@ifxundefined [1]{%
		\@ifx{#1\undefined}
	}%
	\providecommand \@ifnum [1]{%
		\ifnum #1\expandafter \@firstoftwo
		\else \expandafter \@secondoftwo
		\fi
	}%
	\providecommand \@ifx [1]{%
		\ifx #1\expandafter \@firstoftwo
		\else \expandafter \@secondoftwo
		\fi
	}%
	\providecommand \natexlab [1]{#1}%
	\providecommand \enquote  [1]{``#1''}%
	\providecommand \bibnamefont  [1]{#1}%
	\providecommand \bibfnamefont [1]{#1}%
	\providecommand \citenamefont [1]{#1}%
	\providecommand \href@noop [0]{\@secondoftwo}%
	\providecommand \href [0]{\begingroup \@sanitize@url \@href}%
	\providecommand \@href[1]{\@@startlink{#1}\@@href}%
	\providecommand \@@href[1]{\endgroup#1\@@endlink}%
	\providecommand \@sanitize@url [0]{\catcode `\\12\catcode `\$12\catcode
		`\&12\catcode `\#12\catcode `\^12\catcode `\_12\catcode `\%12\relax}%
	\providecommand \@@startlink[1]{}%
	\providecommand \@@endlink[0]{}%
	\providecommand \url  [0]{\begingroup\@sanitize@url \@url }%
	\providecommand \@url [1]{\endgroup\@href {#1}{\urlprefix }}%
	\providecommand \urlprefix  [0]{URL }%
	\providecommand \Eprint [0]{\href }%
	\providecommand \doibase [0]{https://doi.org/}%
	\providecommand \selectlanguage [0]{\@gobble}%
	\providecommand \bibinfo  [0]{\@secondoftwo}%
	\providecommand \bibfield  [0]{\@secondoftwo}%
	\providecommand \translation [1]{[#1]}%
	\providecommand \BibitemOpen [0]{}%
	\providecommand \bibitemStop [0]{}%
	\providecommand \bibitemNoStop [0]{.\EOS\space}%
	\providecommand \EOS [0]{\spacefactor3000\relax}%
	\providecommand \BibitemShut  [1]{\csname bibitem#1\endcsname}%
	\let\auto@bib@innerbib\@empty
	\bibitem [{\citenamefont {Bennett}\ and\ \citenamefont
		{Brassard}(1984)}]{bennett1984quantum}%
	\BibitemOpen
	\bibfield  {author} {\bibinfo {author} {\bibfnamefont {C.~H.}\ \bibnamefont
			{Bennett}}\ and\ \bibinfo {author} {\bibfnamefont {G.}~\bibnamefont
			{Brassard}},\ }\bibfield  {title} {\bibinfo {title} {Quantum cryptography:
			Public key distribution and coin tossing},\ }in\ \href
	{https://www.sciencedirect.com/science/article/pii/S0304397514004241} {\emph
		{\bibinfo {booktitle} {Proceedings of IEEE International Conference on
				Computers, Systems and Signal Processing}}}\ (\bibinfo {organization}
	{Bangalore, India},\ \bibinfo {year} {1984})\ pp.\ \bibinfo {pages}
	{175--179}\BibitemShut {NoStop}%
	\bibitem [{\citenamefont {Shor}(1997)}]{shor1997factoring}%
	\BibitemOpen
	\bibfield  {author} {\bibinfo {author} {\bibfnamefont {P.~W.}\ \bibnamefont
			{Shor}},\ }\bibfield  {title} {\bibinfo {title} {Polynomial-time algorithms
			for prime factorization and discrete logarithms on a quantum computer},\
	}\href {https://doi.org/10.1137/S0097539795293172} {\bibfield  {journal}
		{\bibinfo  {journal} {SIAM Journal on Computing}\ }\textbf {\bibinfo {volume}
			{26}},\ \bibinfo {pages} {1484} (\bibinfo {year} {1997})}\BibitemShut
	{NoStop}%
	\bibitem [{\citenamefont {Wootters}\ and\ \citenamefont
		{Zurek}(1982)}]{Wootters1982clone}%
	\BibitemOpen
	\bibfield  {author} {\bibinfo {author} {\bibfnamefont {W.~K.}\ \bibnamefont
			{Wootters}}\ and\ \bibinfo {author} {\bibfnamefont {W.~H.}\ \bibnamefont
			{Zurek}},\ }\bibfield  {title} {\bibinfo {title} {A single quantum cannot be
			cloned},\ }\href {https://doi.org/10.1038/299802a0} {\bibfield  {journal}
		{\bibinfo  {journal} {Nature}\ }\textbf {\bibinfo {volume} {299}},\ \bibinfo
		{pages} {802} (\bibinfo {year} {1982})}\BibitemShut {NoStop}%
	\bibitem [{\citenamefont {Dieks}(1982)}]{DIEKS1982clone}%
	\BibitemOpen
	\bibfield  {author} {\bibinfo {author} {\bibfnamefont {D.}~\bibnamefont
			{Dieks}},\ }\bibfield  {title} {\bibinfo {title} {Communication by epr
			devices},\ }\href
	{https://doi.org/https://doi.org/10.1016/0375-9601(82)90084-6} {\bibfield
		{journal} {\bibinfo  {journal} {Physics Letters A}\ }\textbf {\bibinfo
			{volume} {92}},\ \bibinfo {pages} {271} (\bibinfo {year} {1982})}\BibitemShut
	{NoStop}%
	\bibitem [{\citenamefont {Barnum}\ \emph {et~al.}(1996)\citenamefont {Barnum},
		\citenamefont {Caves}, \citenamefont {Fuchs}, \citenamefont {Jozsa},\ and\
		\citenamefont {Schumacher}}]{barnum1996broadcasting}%
	\BibitemOpen
	\bibfield  {author} {\bibinfo {author} {\bibfnamefont {H.}~\bibnamefont
			{Barnum}}, \bibinfo {author} {\bibfnamefont {C.~M.}\ \bibnamefont {Caves}},
		\bibinfo {author} {\bibfnamefont {C.~A.}\ \bibnamefont {Fuchs}}, \bibinfo
		{author} {\bibfnamefont {R.}~\bibnamefont {Jozsa}},\ and\ \bibinfo {author}
		{\bibfnamefont {B.}~\bibnamefont {Schumacher}},\ }\bibfield  {title}
	{\bibinfo {title} {Noncommuting mixed states cannot be broadcast},\ }\href
	{https://doi.org/10.1103/PhysRevLett.76.2818} {\bibfield  {journal} {\bibinfo
			{journal} {Phys. Rev. Lett.}\ }\textbf {\bibinfo {volume} {76}},\ \bibinfo
		{pages} {2818} (\bibinfo {year} {1996})}\BibitemShut {NoStop}%
	\bibitem [{\citenamefont {Cotler}\ \emph {et~al.}(2019)\citenamefont {Cotler},
		\citenamefont {Choi}, \citenamefont {Lukin}, \citenamefont {Gharibyan},
		\citenamefont {Grover}, \citenamefont {Tai}, \citenamefont {Rispoli},
		\citenamefont {Schittko}, \citenamefont {Preiss}, \citenamefont {Kaufman},
		\citenamefont {Greiner}, \citenamefont {Pichler},\ and\ \citenamefont
		{Hayden}}]{cotler2019cooling}%
	\BibitemOpen
	\bibfield  {author} {\bibinfo {author} {\bibfnamefont {J.}~\bibnamefont
			{Cotler}}, \bibinfo {author} {\bibfnamefont {S.}~\bibnamefont {Choi}},
		\bibinfo {author} {\bibfnamefont {A.}~\bibnamefont {Lukin}}, \bibinfo
		{author} {\bibfnamefont {H.}~\bibnamefont {Gharibyan}}, \bibinfo {author}
		{\bibfnamefont {T.}~\bibnamefont {Grover}}, \bibinfo {author} {\bibfnamefont
			{M.~E.}\ \bibnamefont {Tai}}, \bibinfo {author} {\bibfnamefont
			{M.}~\bibnamefont {Rispoli}}, \bibinfo {author} {\bibfnamefont
			{R.}~\bibnamefont {Schittko}}, \bibinfo {author} {\bibfnamefont {P.~M.}\
			\bibnamefont {Preiss}}, \bibinfo {author} {\bibfnamefont {A.~M.}\
			\bibnamefont {Kaufman}}, \bibinfo {author} {\bibfnamefont {M.}~\bibnamefont
			{Greiner}}, \bibinfo {author} {\bibfnamefont {H.}~\bibnamefont {Pichler}},\
		and\ \bibinfo {author} {\bibfnamefont {P.}~\bibnamefont {Hayden}},\
	}\bibfield  {title} {\bibinfo {title} {Quantum virtual cooling},\ }\href
	{https://doi.org/10.1103/PhysRevX.9.031013} {\bibfield  {journal} {\bibinfo
			{journal} {Phys. Rev. X}\ }\textbf {\bibinfo {volume} {9}},\ \bibinfo {pages}
		{031013} (\bibinfo {year} {2019})}\BibitemShut {NoStop}%
	\bibitem [{\citenamefont {Temme}\ \emph {et~al.}(2017)\citenamefont {Temme},
		\citenamefont {Bravyi},\ and\ \citenamefont
		{Gambetta}}]{temmeErrorMitigationShortDepth2017}%
	\BibitemOpen
	\bibfield  {author} {\bibinfo {author} {\bibfnamefont {K.}~\bibnamefont
			{Temme}}, \bibinfo {author} {\bibfnamefont {S.}~\bibnamefont {Bravyi}},\ and\
		\bibinfo {author} {\bibfnamefont {J.~M.}\ \bibnamefont {Gambetta}},\
	}\bibfield  {title} {\bibinfo {title} {Error mitigation for short-depth
			quantum circuits},\ }\href {https://doi.org/10.1103/PhysRevLett.119.180509}
	{\bibfield  {journal} {\bibinfo  {journal} {Phys. Rev. Lett.}\ }\textbf
		{\bibinfo {volume} {119}},\ \bibinfo {pages} {180509} (\bibinfo {year}
		{2017})}\BibitemShut {NoStop}%
	\bibitem [{\citenamefont {Endo}\ \emph {et~al.}(2018)\citenamefont {Endo},
		\citenamefont {Benjamin},\ and\ \citenamefont
		{Li}}]{endoPracticalQuantumError2018}%
	\BibitemOpen
	\bibfield  {author} {\bibinfo {author} {\bibfnamefont {S.}~\bibnamefont
			{Endo}}, \bibinfo {author} {\bibfnamefont {S.~C.}\ \bibnamefont {Benjamin}},\
		and\ \bibinfo {author} {\bibfnamefont {Y.}~\bibnamefont {Li}},\ }\bibfield
	{title} {\bibinfo {title} {Practical quantum error mitigation for near-future
			applications},\ }\href {https://doi.org/10.1103/PhysRevX.8.031027} {\bibfield
		{journal} {\bibinfo  {journal} {Phys. Rev. X}\ }\textbf {\bibinfo {volume}
			{8}},\ \bibinfo {pages} {31027} (\bibinfo {year} {2018})}\BibitemShut
	{NoStop}%
	\bibitem [{\citenamefont {Yuan}\ \emph {et~al.}(2024)\citenamefont {Yuan},
		\citenamefont {Regula}, \citenamefont {Takagi},\ and\ \citenamefont
		{Gu}}]{yuan2024virtual}%
	\BibitemOpen
	\bibfield  {author} {\bibinfo {author} {\bibfnamefont {X.}~\bibnamefont
			{Yuan}}, \bibinfo {author} {\bibfnamefont {B.}~\bibnamefont {Regula}},
		\bibinfo {author} {\bibfnamefont {R.}~\bibnamefont {Takagi}},\ and\ \bibinfo
		{author} {\bibfnamefont {M.}~\bibnamefont {Gu}},\ }\bibfield  {title}
	{\bibinfo {title} {Virtual quantum resource distillation},\ }\href
	{https://doi.org/10.1103/PhysRevLett.132.050203} {\bibfield  {journal}
		{\bibinfo  {journal} {Phys. Rev. Lett.}\ }\textbf {\bibinfo {volume} {132}},\
		\bibinfo {pages} {050203} (\bibinfo {year} {2024})}\BibitemShut {NoStop}%
	\bibitem [{\citenamefont {Parzygnat}\ \emph {et~al.}(2024)\citenamefont
		{Parzygnat}, \citenamefont {Fullwood}, \citenamefont {Buscemi},\ and\
		\citenamefont {Chiribella}}]{PhysRevLett.132.110203}%
	\BibitemOpen
	\bibfield  {author} {\bibinfo {author} {\bibfnamefont {A.~J.}\ \bibnamefont
			{Parzygnat}}, \bibinfo {author} {\bibfnamefont {J.}~\bibnamefont {Fullwood}},
		\bibinfo {author} {\bibfnamefont {F.}~\bibnamefont {Buscemi}},\ and\ \bibinfo
		{author} {\bibfnamefont {G.}~\bibnamefont {Chiribella}},\ }\bibfield  {title}
	{\bibinfo {title} {Virtual quantum broadcasting},\ }\href
	{https://doi.org/10.1103/PhysRevLett.132.110203} {\bibfield  {journal}
		{\bibinfo  {journal} {Phys. Rev. Lett.}\ }\textbf {\bibinfo {volume} {132}},\
		\bibinfo {pages} {110203} (\bibinfo {year} {2024})}\BibitemShut {NoStop}%
	\bibitem [{\citenamefont {Yao}\ \emph {et~al.}(2024)\citenamefont {Yao},
		\citenamefont {Liu}, \citenamefont {Zhu},\ and\ \citenamefont
		{Wang}}]{PhysRevA.110.012458}%
	\BibitemOpen
	\bibfield  {author} {\bibinfo {author} {\bibfnamefont {H.}~\bibnamefont
			{Yao}}, \bibinfo {author} {\bibfnamefont {X.}~\bibnamefont {Liu}}, \bibinfo
		{author} {\bibfnamefont {C.}~\bibnamefont {Zhu}},\ and\ \bibinfo {author}
		{\bibfnamefont {X.}~\bibnamefont {Wang}},\ }\bibfield  {title} {\bibinfo
		{title} {Optimal unilocal virtual quantum broadcasting},\ }\href
	{https://doi.org/10.1103/PhysRevA.110.012458} {\bibfield  {journal} {\bibinfo
			{journal} {Phys. Rev. A}\ }\textbf {\bibinfo {volume} {110}},\ \bibinfo
		{pages} {012458} (\bibinfo {year} {2024})}\BibitemShut {NoStop}%
	\bibitem [{\citenamefont {Zheng}\ \emph {et~al.}(2025)\citenamefont {Zheng},
		\citenamefont {Nie}, \citenamefont {Liu}, \citenamefont {Luo}, \citenamefont
		{Lu},\ and\ \citenamefont {Liu}}]{8vrg-tvsd}%
	\BibitemOpen
	\bibfield  {author} {\bibinfo {author} {\bibfnamefont {Y.}~\bibnamefont
			{Zheng}}, \bibinfo {author} {\bibfnamefont {X.}~\bibnamefont {Nie}}, \bibinfo
		{author} {\bibfnamefont {H.}~\bibnamefont {Liu}}, \bibinfo {author}
		{\bibfnamefont {Y.}~\bibnamefont {Luo}}, \bibinfo {author} {\bibfnamefont
			{D.}~\bibnamefont {Lu}},\ and\ \bibinfo {author} {\bibfnamefont
			{X.}~\bibnamefont {Liu}},\ }\bibfield  {title} {\bibinfo {title}
		{Experimental virtual quantum broadcasting},\ }\href
	{https://doi.org/10.1103/8vrg-tvsd} {\bibfield  {journal} {\bibinfo
			{journal} {Phys. Rev. A}\ }\textbf {\bibinfo {volume} {111}},\ \bibinfo
		{pages} {L060402} (\bibinfo {year} {2025})}\BibitemShut {NoStop}%
	\bibitem [{\citenamefont {Fitzsimons}\ \emph
		{et~al.}(2015{\natexlab{a}})\citenamefont {Fitzsimons}, \citenamefont
		{Jones},\ and\ \citenamefont {Vedral}}]{fitzsimons2015quantum}%
	\BibitemOpen
	\bibfield  {author} {\bibinfo {author} {\bibfnamefont {J.~F.}\ \bibnamefont
			{Fitzsimons}}, \bibinfo {author} {\bibfnamefont {J.~A.}\ \bibnamefont
			{Jones}},\ and\ \bibinfo {author} {\bibfnamefont {V.}~\bibnamefont
			{Vedral}},\ }\bibfield  {title} {\bibinfo {title} {Quantum correlations which
			imply causation},\ }\href {https://doi.org/10.1038/srep18281} {\bibfield
		{journal} {\bibinfo  {journal} {Sci. Rep.}\ }\textbf {\bibinfo {volume}
			{5}},\ \bibinfo {pages} {18281} (\bibinfo {year}
		{2015}{\natexlab{a}})}\BibitemShut {NoStop}%
	\bibitem [{\citenamefont {Collins}\ and\ \citenamefont
		{Śniady}(2006)}]{Collins2006random}%
	\BibitemOpen
	\bibfield  {author} {\bibinfo {author} {\bibfnamefont {B.}~\bibnamefont
			{Collins}}\ and\ \bibinfo {author} {\bibfnamefont {P.}~\bibnamefont
			{Śniady}},\ }\bibfield  {title} {\bibinfo {title} {Integration with respect
			to the haar measure on unitary, orthogonal and symplectic group},\ }\href
	{https://doi.org/10.1007/s00220-006-1554-3} {\bibfield  {journal} {\bibinfo
			{journal} {Communications in Mathematical Physics}\ }\textbf {\bibinfo
			{volume} {264}},\ \bibinfo {pages} {773} (\bibinfo {year}
		{2006})}\BibitemShut {NoStop}%
	\bibitem [{\citenamefont {Xiao}\ \emph {et~al.}(2025)\citenamefont {Xiao},
		\citenamefont {Liu},\ and\ \citenamefont {Liu}}]{8g6j-w7ld}%
	\BibitemOpen
	\bibfield  {author} {\bibinfo {author} {\bibfnamefont {Y.}~\bibnamefont
			{Xiao}}, \bibinfo {author} {\bibfnamefont {X.}~\bibnamefont {Liu}},\ and\
		\bibinfo {author} {\bibfnamefont {Z.}~\bibnamefont {Liu}},\ }\bibfield
	{title} {\bibinfo {title} {No practical quantum broadcasting: General
			framework},\ }\href {https://doi.org/10.1103/8g6j-w7ld} {\bibfield  {journal}
		{\bibinfo  {journal} {Phys. Rev. Res.}\ }\textbf {\bibinfo {volume} {7}},\
		\bibinfo {pages} {033194} (\bibinfo {year} {2025})}\BibitemShut {NoStop}%
	\bibitem [{\citenamefont {Fitzsimons}\ \emph
		{et~al.}(2015{\natexlab{b}})\citenamefont {Fitzsimons}, \citenamefont
		{Jones},\ and\ \citenamefont {Vedral}}]{Fitzsimons2015}%
	\BibitemOpen
	\bibfield  {author} {\bibinfo {author} {\bibfnamefont {J.~F.}\ \bibnamefont
			{Fitzsimons}}, \bibinfo {author} {\bibfnamefont {J.~A.}\ \bibnamefont
			{Jones}},\ and\ \bibinfo {author} {\bibfnamefont {V.}~\bibnamefont
			{Vedral}},\ }\bibfield  {title} {\bibinfo {title} {Quantum correlations which
			imply causation},\ }\href {https://doi.org/10.1038/srep18281} {\bibfield
		{journal} {\bibinfo  {journal} {Scientific Reports}\ }\textbf {\bibinfo
			{volume} {5}},\ \bibinfo {pages} {18281} (\bibinfo {year}
		{2015}{\natexlab{b}})}\BibitemShut {NoStop}%
	\bibitem [{\citenamefont {Buscemi}\ \emph {et~al.}(2013)\citenamefont
		{Buscemi}, \citenamefont {Dall'Arno}, \citenamefont {Ozawa},\ and\
		\citenamefont {Vedral}}]{buscemi2013directobservationtwopointquantum}%
	\BibitemOpen
	\bibfield  {author} {\bibinfo {author} {\bibfnamefont {F.}~\bibnamefont
			{Buscemi}}, \bibinfo {author} {\bibfnamefont {M.}~\bibnamefont {Dall'Arno}},
		\bibinfo {author} {\bibfnamefont {M.}~\bibnamefont {Ozawa}},\ and\ \bibinfo
		{author} {\bibfnamefont {V.}~\bibnamefont {Vedral}},\ }\bibfield  {title}
	{\bibinfo {title} {Direct observation of any two-point quantum correlation
			function},\ }\href {https://arxiv.org/abs/1312.4240} {\bibfield  {journal}
		{\bibinfo  {journal} {arXiv preprint arXiv:1312.4240}\ } (\bibinfo {year}
		{2013})}\BibitemShut {NoStop}%
	\bibitem [{\citenamefont {Buscemi}\ \emph {et~al.}(2014)\citenamefont
		{Buscemi}, \citenamefont {Dall'Arno}, \citenamefont {Ozawa},\ and\
		\citenamefont {Vedral}}]{doi:10.1142/S0219749915600023}%
	\BibitemOpen
	\bibfield  {author} {\bibinfo {author} {\bibfnamefont {F.}~\bibnamefont
			{Buscemi}}, \bibinfo {author} {\bibfnamefont {M.}~\bibnamefont {Dall'Arno}},
		\bibinfo {author} {\bibfnamefont {M.}~\bibnamefont {Ozawa}},\ and\ \bibinfo
		{author} {\bibfnamefont {V.}~\bibnamefont {Vedral}},\ }\bibfield  {title}
	{\bibinfo {title} {Universal optimal quantum correlator},\ }\href
	{https://doi.org/10.1142/S0219749915600023} {\bibfield  {journal} {\bibinfo
			{journal} {International Journal of Quantum Information}\ }\textbf {\bibinfo
			{volume} {12}},\ \bibinfo {pages} {1560002} (\bibinfo {year}
		{2014})}\BibitemShut {NoStop}%
	\bibitem [{\citenamefont {Horsman}\ \emph {et~al.}(2017)\citenamefont
		{Horsman}, \citenamefont {Heunen}, \citenamefont {Pusey}, \citenamefont
		{Barrett},\ and\ \citenamefont {Spekkens}}]{horsman2017qsot}%
	\BibitemOpen
	\bibfield  {author} {\bibinfo {author} {\bibfnamefont {D.}~\bibnamefont
			{Horsman}}, \bibinfo {author} {\bibfnamefont {C.}~\bibnamefont {Heunen}},
		\bibinfo {author} {\bibfnamefont {M.~F.}\ \bibnamefont {Pusey}}, \bibinfo
		{author} {\bibfnamefont {J.}~\bibnamefont {Barrett}},\ and\ \bibinfo {author}
		{\bibfnamefont {R.~W.}\ \bibnamefont {Spekkens}},\ }\bibfield  {title}
	{\bibinfo {title} {Can a quantum state over time resemble a quantum state at
			a single time?},\ }\href {https://doi.org/10.1098/rspa.2017.0395} {\bibfield
		{journal} {\bibinfo  {journal} {Proceedings of the Royal Society A:
				Mathematical, Physical and Engineering Sciences}\ }\textbf {\bibinfo {volume}
			{473}},\ \bibinfo {pages} {20170395} (\bibinfo {year} {2017})}\BibitemShut
	{NoStop}%
	\bibitem [{\citenamefont {Fullwood}\ and\ \citenamefont
		{Parzygnat}(2022)}]{fullwood2022qsot}%
	\BibitemOpen
	\bibfield  {author} {\bibinfo {author} {\bibfnamefont {J.}~\bibnamefont
			{Fullwood}}\ and\ \bibinfo {author} {\bibfnamefont {A.~J.}\ \bibnamefont
			{Parzygnat}},\ }\bibfield  {title} {\bibinfo {title} {On quantum states over
			time},\ }\href {https://doi.org/10.1098/rspa.2022.0104} {\bibfield  {journal}
		{\bibinfo  {journal} {Proceedings of the Royal Society A: Mathematical,
				Physical and Engineering Sciences}\ }\textbf {\bibinfo {volume} {478}},\
		\bibinfo {pages} {20220104} (\bibinfo {year} {2022})}\BibitemShut {NoStop}%
	\bibitem [{\citenamefont {Parzygnat}\ and\ \citenamefont
		{Fullwood}(2023)}]{PRXQuantum.4.020334}%
	\BibitemOpen
	\bibfield  {author} {\bibinfo {author} {\bibfnamefont {A.~J.}\ \bibnamefont
			{Parzygnat}}\ and\ \bibinfo {author} {\bibfnamefont {J.}~\bibnamefont
			{Fullwood}},\ }\bibfield  {title} {\bibinfo {title} {From time-reversal
			symmetry to quantum bayes' rules},\ }\href
	{https://doi.org/10.1103/PRXQuantum.4.020334} {\bibfield  {journal} {\bibinfo
			{journal} {PRX Quantum}\ }\textbf {\bibinfo {volume} {4}},\ \bibinfo {pages}
		{020334} (\bibinfo {year} {2023})}\BibitemShut {NoStop}%
	\bibitem [{\citenamefont {Lie}\ and\ \citenamefont
		{Ng}(2024)}]{PhysRevResearch.6.033144}%
	\BibitemOpen
	\bibfield  {author} {\bibinfo {author} {\bibfnamefont {S.~H.}\ \bibnamefont
			{Lie}}\ and\ \bibinfo {author} {\bibfnamefont {N.~H.~Y.}\ \bibnamefont
			{Ng}},\ }\bibfield  {title} {\bibinfo {title} {Quantum state over time is
			unique},\ }\href {https://doi.org/10.1103/PhysRevResearch.6.033144}
	{\bibfield  {journal} {\bibinfo  {journal} {Phys. Rev. Res.}\ }\textbf
		{\bibinfo {volume} {6}},\ \bibinfo {pages} {033144} (\bibinfo {year}
		{2024})}\BibitemShut {NoStop}%
	\bibitem [{\citenamefont {Lie}\ and\ \citenamefont
		{Fullwood}(2024)}]{lie2024uniquemultipartiteextensionquantum}%
	\BibitemOpen
	\bibfield  {author} {\bibinfo {author} {\bibfnamefont {S.~H.}\ \bibnamefont
			{Lie}}\ and\ \bibinfo {author} {\bibfnamefont {J.}~\bibnamefont {Fullwood}},\
	}\bibfield  {title} {\bibinfo {title} {Unique multipartite extension of
			quantum states over time},\ }\href {https://arxiv.org/abs/2410.22630}
	{\bibfield  {journal} {\bibinfo  {journal} {arXiv preprint arXiv:2410.22630}\
		} (\bibinfo {year} {2024})}\BibitemShut {NoStop}%
	\bibitem [{\citenamefont {Liu}\ \emph {et~al.}(2025)\citenamefont {Liu},
		\citenamefont {Qiu}, \citenamefont {Dahlsten},\ and\ \citenamefont
		{Vedral}}]{liu2023quantum}%
	\BibitemOpen
	\bibfield  {author} {\bibinfo {author} {\bibfnamefont {X.}~\bibnamefont
			{Liu}}, \bibinfo {author} {\bibfnamefont {Y.}~\bibnamefont {Qiu}}, \bibinfo
		{author} {\bibfnamefont {O.}~\bibnamefont {Dahlsten}},\ and\ \bibinfo
		{author} {\bibfnamefont {V.}~\bibnamefont {Vedral}},\ }\bibfield  {title}
	{\bibinfo {title} {Quantum causal inference with extremely light touch},\
	}\href {https://doi.org/10.1038/s41534-024-00956-0} {\bibfield  {journal}
		{\bibinfo  {journal} {npj Quantum Inf.}\ }\textbf {\bibinfo {volume} {11}}
		(\bibinfo {year} {2025})}\BibitemShut {NoStop}%
	\bibitem [{\citenamefont {Hoeffding}(1963)}]{Hoeffding01031963}%
	\BibitemOpen
	\bibfield  {author} {\bibinfo {author} {\bibfnamefont {W.}~\bibnamefont
			{Hoeffding}},\ }\bibfield  {title} {\bibinfo {title} {Probability
			inequalities for sums of bounded random variables},\ }\href
	{https://www.tandfonline.com/doi/abs/10.1080/01621459.1963.10500830}
	{\bibfield  {journal} {\bibinfo  {journal} {Journal of the American
				Statistical Association}\ }\textbf {\bibinfo {volume} {58}},\ \bibinfo
		{pages} {13} (\bibinfo {year} {1963})}\BibitemShut {NoStop}%
	\bibitem [{\citenamefont {Vandenberghe}\ and\ \citenamefont
		{Boyd}(1996)}]{doi:10.1137/1038003}%
	\BibitemOpen
	\bibfield  {author} {\bibinfo {author} {\bibfnamefont {L.}~\bibnamefont
			{Vandenberghe}}\ and\ \bibinfo {author} {\bibfnamefont {S.}~\bibnamefont
			{Boyd}},\ }\bibfield  {title} {\bibinfo {title} {Semidefinite programming},\
	}\href {https://doi.org/10.1137/1038003} {\bibfield  {journal} {\bibinfo
			{journal} {SIAM Review}\ }\textbf {\bibinfo {volume} {38}},\ \bibinfo {pages}
		{49} (\bibinfo {year} {1996})}\BibitemShut {NoStop}%
	\bibitem [{\citenamefont {Choi}(1975)}]{CHOI1975285}%
	\BibitemOpen
	\bibfield  {author} {\bibinfo {author} {\bibfnamefont {M.-D.}\ \bibnamefont
			{Choi}},\ }\bibfield  {title} {\bibinfo {title} {Completely positive linear
			maps on complex matrices},\ }\href
	{https://doi.org/https://doi.org/10.1016/0024-3795(75)90075-0} {\bibfield
		{journal} {\bibinfo  {journal} {Linear Algebra Appl}\ }\textbf {\bibinfo
			{volume} {10}},\ \bibinfo {pages} {285} (\bibinfo {year} {1975})}\BibitemShut
	{NoStop}%
	\bibitem [{\citenamefont {Jamio{\l}kowski}(1972)}]{JAMIOLKOWSKI1972275}%
	\BibitemOpen
	\bibfield  {author} {\bibinfo {author} {\bibfnamefont {A.}~\bibnamefont
			{Jamio{\l}kowski}},\ }\bibfield  {title} {\bibinfo {title} {Linear
			transformations which preserve trace and positive semidefiniteness of
			operators},\ }\href
	{https://doi.org/https://doi.org/10.1016/0034-4877(72)90011-0} {\bibfield
		{journal} {\bibinfo  {journal} {Rep. Math. Phys.}\ }\textbf {\bibinfo
			{volume} {3}},\ \bibinfo {pages} {275} (\bibinfo {year} {1972})}\BibitemShut
	{NoStop}%
\end{thebibliography}
%


\end{document}